\documentclass[aps,prd,twocolumn,superscriptaddress,floatfix,showpacs]{revtex4}
\usepackage{graphics}
\usepackage{graphicx}
\usepackage{epsfig}
\usepackage{amsmath}
\usepackage{amsfonts}
\usepackage{bm}
\usepackage{url}

\def\beq{\begin{equation}}
\def\eeq{\end{equation}}
\def\bea{\begin{eqnarray}}
\def\eea{\end{eqnarray}}

\def\k{\kappa}
\def\n{\nu}
\def\lk{\lambda^{\k}}
\def\lnu{\lambda^{\n}}
\def\ms{M_{\odot}}

\def\prd{\reff@jnl{Phys. Rev. D }}        % Physical Review D
\def\cqg{\reff@jnl{Class. Quantum Grav. }} % Classical and Quantum Gravity

\begin{document}
%% \normalsize  
\input epsf.tex

\title{Fisher vs. Bayes : A comparison of parameter estimation techniques for massive black hole binaries to high redshifts with eLISA.}

\author{Edward K. \surname{Porter}}
\email[]{porter@apc.univ-paris7.fr}
\affiliation{Fran\c{c}ois Arago Centre, APC, Universit\'e Paris Diderot,\\ CNRS/IN2P3, CEA/Irfu, Observatoire de Paris, Sorbonne Paris Cit\'e, \\10 rue A. Domon et L. Duquet, 75205 Paris Cedex 13, France}

\author{Neil J. \surname{Cornish}}
\email[]{cornish@physics.montana.edu}
\affiliation{Department of Physics, Montana State University, Bozeman, MT 59717, USA}

\vspace{1cm}
\begin{abstract}
Massive black hole binaries are the primary source of gravitational waves (GW) for the future eLISA observatory.  The detection and parameter estimation
of these sources to high redshift would provide invaluable information on the formation mechanisms of seed black holes, and on the evolution of massive
black holes and their host galaxies through cosmic time.  The Fisher information matrix has been the standard tool for GW parameter estimation in the last
two decades.  However, recent studies have questioned the validity of using the Fisher matrix approach. For example, the Fisher matrix approach sometimes
predicts errors of $\geq100\%$ in the estimation of parameters such as the luminosity distance and sky position.  With advances in computing power, Bayesian inference
is beginning to replace the Fisher matrix approximation in parameter estimation studies. In this work, we 
conduct a Bayesian inference analysis for 120 sources situated at redshifts of between $0.1\leq z\leq 13.2$, and compare the results with those from a Fisher
matrix analysis.  The Fisher matrix results suggest that for this particular selection of sources, eLISA would be unable to localize sources at redshifts of $z\lesssim6$.
In contrast, Bayesian inference provides finite error estimations for all sources in the study, and shows that we can establish minimum closest distances for all sources.
The study further predicts that we should be capable with eLISA, out to a redshift of at least $z\leq13$, of predicting a maximum error in the  
chirp mass of $\lesssim 1\%$, the reduced mass of $\lesssim20\%$, the time to coalescence of 
2 hours, and to a redshift of $z\sim5$, the inclination of the source with a maximum error of $\sim60$ degrees. 
\\

\end{abstract}

\maketitle

%%%%%%%%%%%%%%%%%%%%%%%%%%%%%%%%%%%%%%%%%%%%%%%%%%%%%%%%%%%%%%%%%%%%%%%

\section{Introduction}
Through observations of nearby galaxies, it is appears that a supermassive black hole (SMBH) lies at the center of most, if not every, galaxy.  Evidence
further suggests that there is an intricate connection between the mass and evolution of the SMBH and the evolution of the host galaxy~\cite{Ferrarese:2000se,2000ApJ...539L..13G,2002ApJ...574..740T}.  One of the major unsolved
problems in astrophysics is how do the SMBHs form, and how do they affect galactic evolution over cosmic time?  There are two main theories as to how seed black holes
may have formed in the distant universe.  In the first scenario, the first short-lived, low metallicity, stars produced low mass remnants at redshifts of $z\sim20$.  These so-called
Pop III stars would have produced black hole seeds with masses of a few tens to a few thousand solar masses~\cite{1996ApJ...464..523H,2003ApJ...591..288H,1997ApJ...474....1T,2001ApJ...551L..27M,2003ApJ...596...34B,2011ApJ...727..110C,Bernadetta:2008bc,2012ApJ...745...50W}.  In the second scenario, the direct collapse of the 
centers of protogalactic disks would have produced black holes with masses of $10^5-10^6\,\ms$ at redshifts of $z\lesssim12$~\cite{2006MNRAS.370..289B,2006MNRAS.371.1813L,2012ApJ...749...37M,2010MNRAS.409.1057D}.  As of yet, there is no 
observational evidence to support one model over the other  However, we do know that from observations of high redshift quasars, that SMBHs with masses of $10^9\,\ms$ existed at
$z\approx7$~\cite{2011Natur.474..616M}.

It is also now accepted that whatever the mechanism of seed formation, the most likely explanation for the masses of SMBHs that we observe in the local universe is a model
of hierarchical structure formation through mergers~\cite{Volonteri:2002vz,Volonteri:2012tp,2010ApJ...719..851S,2014ApJ...794..104S,2012MNRAS.423.2533B,2014MNRAS.442.3616L}.  While observational evidence strongly supports galactic mergers, the fate of the central SMBHs is unclear.  It is believed that after the 
galactic merger has taken place, the two SMBHs sink to the bottom of the common potential well due to dynamical friction with the galactic environment~\cite{Ostriker:1998fa,Milosavljevic:2001vi}.  The two black
holes then become gravitationally bound and form a binary (SMBHB) when the separation is on the order of $\leq10$ parsecs~\cite{Dotti:2006ef}.  The subsequent evolution to the point where gravitational
waves (GW) become the dominant mechanism of energy loss in the binary is still unknown, but is probably dependent on the role of mechanisms such as gas dynamics, the spins 
of the black holes, the stellar field near the galactic center and the possible existence of circum-black hole or circum-binary accretion disks.

To compound the uncertainty in the final evolution of the SMBHB, there is little observational evidence for the existence of SMBHBs (see ~\cite{Bogdanovic:2014cua} for an interesting review).  Among
the current strongest candidates are: the radio galaxy 0402+379~\cite{2006ApJ...646...49R} which has an estimated total mass of $1.5\times10^8\,\ms$.  The separation between the two SMBHs is
estimated to be 7.3 pc, giving an orbital period of 150,000 years, and the distance to the source is $z=0.055$.  OJ 287 is purported to contain two SMBHs with individual masses
of $(1.8\times10^9,10^8)\,\ms$, and is at a redshift of $z=0.306$~\cite{Valtonen:2008aj}.  It is believed that the smaller black hole orbits the larger object with an orbital period of 11-12 years.  Finally,
a potential sub-milliparsec binary was discovered in the galaxy SDSS J1201+30, at a distance of $z=0.146$~\cite{2014ApJ...786..103L}.  Unfortunately, all of these candidates are at too low a redshift
to usefully say anything about the high-$z$ behaviour of SMBHBs.  Furthermore, the errors in the mass estimates are on the scale of orders of magnitude, making it difficult to 
say anything useful on the black hole mass function in the universe.

To try and answer some of these questions, ESA has recently chosen the theme of the ``Gravitational Wave Universe" for the Cosmic Vision L3 mission selection.  Within this 
program, a GW observatory called eLISA has been proposed~\cite{Whitepaper,ngoscience}.  As the observatory will function in the frequency band of $10^{-5}\leq f/Hz \leq 1$, one of
 the main source targets will be the merger of SMBHBs to high redshift.  We expect that with eLISA, it should be possible to measure the system parameters with sufficient accuracy that one could investigate the models of BH seed formation~\cite{Gair:2010bx,Sesana:2010wy}.

In the last decade there has been a lot of progress made in the development of advanced algorithms that search the parameter space for SMBHBs.  In general, because the
signal-to-noise ratio (SNR) for these sources is so high, the actual detection is relatively simple.  It is the estimation of the system parameters however that is the most time
consuming.  Until recently, the Fisher Information matrix (FIM) was the standard parameter estimation method in GW astronomy.  The reason for this was two-fold : firstly, it was known 
from information theory that the inverse of the FIM provided the variance in the estimation of parameters.  The fact that this calculation could be done quickly (i.e. seconds to
minutes), and with a relatively small investment in coding, made a FIM a popular choice.  Secondly, in the late 1980s, the emerging field of information geometry demonstrated 
that the FIM could be associated with a metric tensor on the parameter manifold, and to the local curvature of the log-likelihood.  At a time when template banks were the only
conceivable option to search for GWs, the FIM was implemented in the proper spacing of templates in the bank~\cite{finn1992,Owen:1995tm,Owen:1998dk,Porter:2002vk}.  The early success of the FIM ensured that it stayed as the
main parameter estimation tool within the GW community. 

In the last decade however, and mostly due to the leap in computational power, there has been a paradigm shift within the community when it comes to parameter estimation.  It
was demonstrated very early on that template banks would be computationally prohibitive for a space-based observatory~\cite{Gair:2004iv,Cornish:2005hd,Cornish2006ms}.  This pushed the space-based GW community
to move to more dynamic Bayesian based algorithms.  For the question of parameter estimation, different variants of a Markov Chain Monte Carlo (MCMC) became very 
popular~\cite{Cornish2006ry,Cornish2006dt,Cornish2006ms,Cornish:2005qw,Porter:2013wwa,vecchiowickham2004,Stroeer:2006ye,Stroeer:2007tg,Rover:2007iq,Veitch:2008ur,Trias:2008dc,Cornish:2007if,Littenberg:2009bm}.  While the method required a much higher investment in implementation of the algorithm, and while it took much longer to run (hours to days), the benefit was that
one obtained the marginalized posterior distributions for each of the parameters and allowed for the easy incorporation of prior information.  If the algorithm had converged, then one had all the information that one could ask for regarding the problem at hand.

However, what began as an effort to implement an alternative method that would confirm and solidify the results of the FIM, we very quickly began to observe situations where
the results from the two methods did not agree.  On further investigation, we began to see many problems with the FIM calculation (e.g. 100\% errors in the estimation of the
luminosity distance, or errors in the sky position larger than the entire sky).  For the first time, the community began to question the validity of using the FIM (see, for example,
Ref~\cite{Vallisneri:2007ev} for an in-depth analysis).  As time has progressed, confidence in the reliability of Bayesian inference methods have grown, while the FIM is now viewed as
an approximation method to be used with caution.

All of this has led to an unfortunate, and unpalatable situation, where we are now using advanced Bayesian inference methods to demonstrate parameter estimation capabilities 
for individual sources, of many different source types, but revert to the FIM when large scale parameter estimation studies need to be conducted, even though we know a-priori that
many of the results will be incorrect.  Unfortunately, these types of studies are usually associated with science impact investigations for mission formulation, upon which policy decisions
are made (see, for example~\cite{nasa2012,ngoscience}).  As, to our knowledge, this is the largest ever Bayesian inference study undertaken for SMBHBs,  the goal of this work is to take the first step towards large-scale Bayesian inference studies,
and to caution against drawing conclusion from a FIM study.

%%%%%%%%%%%%%%%%%%%%%%%%%%%%%%%%%%%%%%%%%%%%%%%%%%%%%%%%%%%%%%%%%%%%%%%

\subsection{Outline of the paper}
The paper is structured as follows.  In Sec~\ref{sec:elisa} we define the eLISA mission concept, and outline the response of the observatory to an impinging GW.  We also
define the higher harmonic inspiral waveform that will be used for the study.  Sec.~\ref{sec:statmeth} introduces the Fisher information matrix  and discusses potential failings of the
method.  This is followed by a description of the MCMC algorithm implemented in this work.  In Sec.~\ref{sec:astro} we discuss the generation of the sources that were studied,  as 
well as the parameter priors used for the Bayesian inference.  Sec.~\ref{sec:results} presents the main results comparing the Fisher matrix and Bayesian inference methods. 

%%%%%%%%%%%%%%%%%%%%%%%%%%%%%%%%%%%%%%%%%%%%%%%%%%%%%%%%%%%%%%%%%%%%%%%

\section{The \lowercase{e}LISA Detector Response}\label{sec:elisa}
\begin{figure}[t!]
 \includegraphics[width=\columnwidth, height=2.5in]{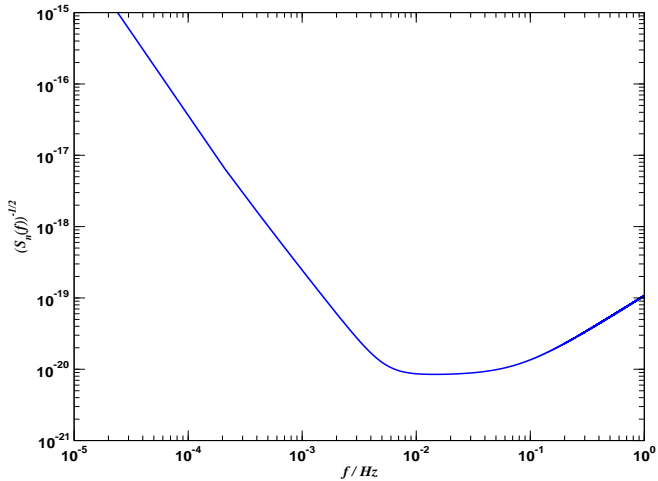}
 \caption{\label{fig:noise}Instrumental noise model for a $10^9$m arm eLISA configuration.}
 \label{fig:psd}
\end{figure}

The European Space Agency (ESA) has recently chosen the theme of the ``Gravitational Wave Universe" for the L3 mission within the Cosmic Vision framework.
For this theme, a mission concept called eLISA has been proposed. This mission is a restructured version of the LISA mission, composed again of a constellation of three satellites, following ballistic heliocentric orbits.  The main differences between eLISA and LISA are (i) a shortening of the arm-length between spacecraft from five to one millions kilometers, and (ii) four, rather than six, laser links.  The idea behind the mission is to launch the three satellites into an orbit roughly $20^{o}$ behind the earth, inclined at $60^{o}$ to the ecliptic.  The constellation is composed of one ``mother" spacecraft at the center of the constellation, with two outlier ``daughter" spacecraft.  The constellation retains a $60^{o}$ angle between spacecraft, with a  nominal mission lifetime of two years.  In Figure~\ref{fig:psd} we present the eLISA instrumental noise curve.   Compared to the LISA noise curve, we are worse off in sensitivity by a factor of 5 at lower frequencies, and the ``bucket" of the noise
curve has moved upwards and two the right.  This makes the constellation more sensitive to lower mass binaries.  At the highest frequencies, we gain a little 
sensitivity due to a reduction of photon shot noise, coming from the shorter armlengths.

In the low frequency approximation~\cite{Cutler:1997ta}, where the GW wavelength is greater than the arm length of the detector, the strain of the gravitational wave (GW) in the eLISA detector with both polarizations is given by the linear combination
\begin{equation}
h(t) = h_{+}(\xi(t))F^{+}+h_{\times}(\xi(t))F^{\times}.
\end{equation}
Here we define the phase shifted time parameter $\xi(t)$ as 
\begin{equation}
\xi(t) = t - R_{\oplus}\sin\theta\cos\left(\alpha(t) - \phi\right),
\end{equation}
where $R_{\oplus} = 1 AU / c$ is the radial distance to the detector guiding center, $c$ is the speed of light, $\left(\theta,\phi\right)$ are the position angles of the source in the sky, $\alpha(t)=2\pi f_{m}t + \kappa$, $f_{m}=1/year$ is the LISA modulation frequency and $\kappa$ gives the initial ecliptic longitude of the guiding center.  Due to the shorter arm length of eLISA, the LFA is a good approximation for the detector response up to a frequency of 0.2 Hz.  It was demonstrated that the inclusion of higher harmonic corrections into the
comparable mass waveform dramatically improves parameter estimation and reduces correlations between parameters~\cite{Moore1999zw,Sintes1999cg,Porter2008kn,Trias2007fp,Arun2007hu}.  It is clear that with reduced sensitivity at low frequencies, the higher harmonic corrections will be 
indispensible for detecting and resolving the most massive binaries.  With this in mind, the post-Newtonian (PN) inspiral 
waveform polarizations, including higher harmonic corrections up to 2-PN order are defined by~\cite{Blanchet1996pi}
\begin{eqnarray}\label{eqn:strain}
 h_{+,\times} & = & \frac{2 G M \eta}{c^2 D_L} x\left[ H^{(0)}_{+,\times} + x^{1/2} H^{(1/2)}_{+,\times} \right. \nonumber \\
              &  + &  \left.  x H^{(1)}_{+,\times} + x^{3/2} H^{(3/2)}_{+,\times} + x^2 H^{(2)}_{+,\times} \right], \\
              \nonumber      
\end{eqnarray}
Here $m=m_{1}+m_{2}$ is the total mass of the binary, $\eta = m_{1}m_{2}/m^{2}$ is the reduced mass ratio and $D_{L}$ is the luminosity distance of the source.  We can define a relation between redshift, $z$, and luminosity distance, $D_{L}$, within a $\Lambda$CDM model by
\begin{equation}
 D_L = (1+z) \frac{c}{H_0} \int_0^z  \frac{dz'}{\sqrt{\Omega_R (1+z')^4 + \Omega_M (1+z')^3 + \Omega_\Lambda}},
\end{equation}
where we use the concurrent PLANCK values of $\Omega_{R}=4.9\times10^{-5}$, $\Omega_{M} = 0.3086$ and $\Omega_{\Lambda} = 0.6914$ and a Hubble's constant of $H_{0}=67$ km/s/Mpc~\cite{Planck2013nga}.
The invariant PN velocity parameter is defined by, $x = \left(Gm\omega / c^{3}\right)^{2/3}$, where $\omega=d\Phi_{0rb}/dt$ is the 2 PN order circular orbital frequency and $\Phi_{orb}=1/2\Phi_{GW}=\varphi_{c}^{orb}-\phi_{orb}(t)$ is the orbital phase defined by
\begin{eqnarray}
\Phi_{orb}(t) &=& \varphi_{c}^{orb}-\frac{1}{\eta}\left[\Theta^{5/8}+\left(\frac{3715}{8064}+\frac{55}{96}\eta\right)\Theta^{3/8}-\frac{3\pi}{4}\Theta^{1/4}\right.\nonumber\\
 &+&\left.\left(\frac{9275495}{14450688}+\frac{284875}{258048}\eta+\frac{1855}{2048}\eta^{2}\right)\Theta^{1/8}\right],
\label{eqn:phase}
\end{eqnarray}
where $\Theta$ is related to the time to coalescence of the wave, $t_{c}$, by
\begin{equation}
\Theta(t) = \frac{c^{3}\eta}{5Gm}\left(t_{c}-t\right),
\end{equation}
and $\varphi_{c}^{orb}$ is the orbital phase of the wave at coalescence.  For the rest of the paper, we will work with GW phases.  In Eqn~(\ref{eqn:strain}), the functions $H_{+,\times}^{(n)}$ contain the PN corrections to the amplitude and the phase harmonics.  All of the half-integer, and some parts of the higher integer $H_{+,\times}^{(n)}$ terms contain the factor $\delta m = m_1 - m_2$ which supress the odd phase harmonics and parts of the even phase harmonics in the equal mass case.  Furthermore, we see the extra frequency harmonics arising due to the higher order $x^n$ terms, and the fact that the different  $H_{+,\times}^{(n)}$ terms contain different multiples of $n\Phi_{orb}$.

The functions $F^{+,\times}$ are the beam pattern functions of the detector, given in the low frequency approximation by 
\begin{eqnarray}
 F^+_k(t; \psi, \theta, \phi) & = & \frac{1}{2} \left[ \cos(2\psi) D^+(t; \psi, \theta, \phi, \lambda_k) \right. \nonumber \\
                              &  - & \left.  \sin(2\psi) D^\times(t; \psi, \theta, \phi, \lambda_k) \right], \\
 F^\times_k(t; \psi, \theta, \phi) & = & \frac{1}{2} \left[ \sin(2\psi) D^+(t; \psi, \theta, \phi, \lambda_k) \right. \nonumber\\
                              & +  & \left.  \cos(2\psi) D^\times(t; \psi, \theta, \phi, \lambda_k) \right], \\
 \nonumber
 \end{eqnarray}
where $\psi$ is the polarization angle of the wave and we take $\lambda = 0$ for the single channel eLISA configuration.  Explicit expression for the detector pattern functions $D^{+,\times}(t)$ can be found in~\cite{rubbocornish2004}.
The above equations for the frequency and phase can sometimes break down before we reach the last stable circular orbit (LSO) at $R=6M$.  To compensate for this, we terminate the waveforms at $R=7M$ in the cases where the coalescence time is shorter that the duration of the signal,  $t_c < T_{obs}$.  For low mass sources where the LSO frequency is possibly close to the limit or outside the band of the detector, we terminate the waveform when the frequency of the highest harmonic reaches 0.2 Hz.  
%%%%%%%%%%%%%%%%%%%%%%%%%%%%%%%%%%%%%%%%%%%%%%%%%%%%%%%%%%%%%%%%%%%%%%%

\section{Statistical methods for estimating parameter errors.}\label{sec:statmeth}

%%%%%%%%%%%%%%%%%%%%%%%%%%%%%%%%%%%%%%%%%%%%%%%%%%%%%%%%%%%%%%%%%%%%%%%

\subsection{The Fisher information matrix.}\label{sec:pe}
One of the main tools used in the GW community for the estimation of parameter errors is the Fisher information matrix (FIM).  In the high SNR limit, assuming that the
 waveform is a linear function of the waveform parameters, and the error distributions in the parameter errors are Gaussian, the inverse of the FIM is said to give the variance-covariance matrix.  The square root of the diagonal elements of the inverse FIM thus give a 1-$\sigma$ estimation of the error in our parameter estimation.    Given a template $h(t;\lk)$, the FIM is defined by
\beq
\Gamma_{\k\nu} = \left<\frac{\partial h}{\partial \lk}\left|\frac{\partial h}{\partial \lnu}\right. \right> = -E\left[ \frac{\partial^2 \ln {\mathcal L}}{\partial \lk \partial \lnu}\right],
\label{eqn:FIM}
\eeq
where $h = \tilde{h}(f;\lk)$ is the waveform template in the Fourier domain and the angular brackets denote the Fourier domain noise-weighted inner product
\begin{equation}
\left<h\left|s\right.\right> =2\int_{0}^{\infty}\frac{df}{S_{n}(f)}\,\left[ \tilde{h}(f)\tilde{s}^{*}(f) +  \tilde{h}^{*}(f)\tilde{s}(f) \right],
\label{eqn:scalarprod}
\end{equation}
where a tilde denotes a Fourier transform and an asterisk denotes a complex conjugate.  The quantity $S_{n}(f)$ is the one-sided noise spectral density of the detector, which is normally a combination of instrumental and galactic confusion noise, but in this article represents the instrumental noise model for eLISA presented above.  Finally, the 
FIM can also be interpreted as the second derivative of the log-likelihood, $\ln {\mathcal L}$, where we define the likelihood given a detector output $s(t)$ and a GW
template $h(t)$ as
\beq
 {\mathcal L}(\lk) = \exp\left(-\frac{1}{2}\left<s-h(\lk)\left|s-h(\lk)\right.\right>\right).
 \label{eqn:likelihood}
\eeq

 For almost two decades, the FIM has been the primary tool within the GW community when it comes to parameter estimation.  It has mostly gained popularity from
 studies on the construction of template banks as it was shown that there is a close relationship between the FIM and the metric tensor on the parameter search space~\cite{Owen:1995tm,Owen:1998dk,Porter:2002vk}.  This
 dual interpretation of the FIM as both a statistical and geometric tool have proved to be very useful.  A further advantage of the FIM is the speed of calculation.  In 
 general, one could calculate the FIM in seconds to minutes, depending on the mass of the system and the duration of the signal.  This ability to obtain a quick
 error estimate for the system parameters concretized the usefulness of the FIM for GW astronomy.
 
 However, the FIM has not been without its difficulties.  The first, and probably the most influential, is the interpretation of the FIM results.  This depends heavily on whether
 the problem in hand is attacked from a frequentist or Bayesian approach.  Ref~\cite{Vallisneri:2007ev} has very nicely laid out the main ways of interpreting the FIM, which
 also clearly displays the confusion arising in its interpretation : (i) $\Gamma_{\k\nu}^{-1}$ is a lower bound (Cram\'er-Rao) for the error covariance of any
 unbiased estimator of the true parameter values.  It is therefore a frequentist error for an experiment characterized by a true signal and a particular noise
 realization. (ii) $\Gamma_{\k\nu}^{-1}$ is the frequentist error in the maximum likelihood estimator for the true parameter, assuming Gaussian noise and a high
 signal to noise ratio (SNR). (iii) $\Gamma_{\k\nu}^{-1}$ gives the covariance of the posterior probability distribution for the true parameters, inferred from a 
 single experiment, assuming the presence of a true signal, Gaussian noise and a high SNR.  In this respect, the FIM provides a level of Bayesian uncertainty rather than an error.
 
 As most GW transient sources fall into the final category and require a Bayesian interpretation, one would normally use interpretation (iii) for the FIM.  However, this
 interpretation also has a number of other problems.  Firstly, as can be seen from Eq.~\ref{eqn:FIM}, the FIM is a pure template calculation which assumes a certain
 model for $S_n(f)$.  The actual detector data is not used for the calculation.  As we can also see from the right hand side of the same equation, the FIM can be further interpreted as the expectation value for the local
 curvature of the log-likelihood.  In this respect, the FIM has no knowledge of other possible modes of the solution or on the general structure of the likelihood
 surface.  Finally, even if the SNR is large, $\Gamma_{\k\nu}$ can be either singular, or at least ill conditioned (which happens quite frequently in GW astronomy),
 which results in problems as $\Gamma_{\k\nu}$ has in general to be inverted numerically.  The only possible solution in this case is then a more robust Monte Carlo study. 
 
Another concern with the FIM comes from the very statistics-geometry duality that established its place in the GW community in the first place.  Geometrically, the FIM can be interpreted as being related to both the 
local metric tensor on the parameter space, and as a local estimate of the curvature of the log likelihood function.  Crucially however, the FIM $\Gamma_{\k\nu}$ is not 
a coordinate invariant quantity.  This immediately poses the problem of what is the best choice of coordinates for a particular problem.  For example, it was demonstrated in Ref~\cite{Adams:2012qw} that one could approximate the true likelihood for some sources using the FIM.  It was further 
demonstrated that using a FIM with a $u=1/D_L$ parameterization provides a much closer approximation to the
true likelihood than a parameterization in $D_L$ itself.   Furthermore, as stated earlier, one of the primary conditions for using the FIM in a statistical analysis is
that the distributions in parameter errors are Gaussian.  In reality, data in physics, astrophysics, economics etc. is rarely normal.  In this and other studies, we have
observed a number of situations where the posterior distribution in $D_L$, for example, follows a log-normal distribution.  This suggests that for these sources, an
error estimate from a FIM using a $\ln D_L$ parameterization will be more exact than an estimation using $D_L$.  However, this information is obtained via an a-posteriori
 comparison with another method, so no global statement can be made on the a-priori choice of parameterization for a problem.   We should also again highlight
the fact that in GW astronomy $\Gamma_{\k\nu}$ is almost always ill-conditioned.  In this case, working with the bare physical parameters returns a condition number
for $\Gamma_{\k\nu}$ that effectively prevents a numerical inversion of the matrix.  However, working in logarithms of the very small/large valued parameters 
sufficiently reduces the condition number such that in most, but not all cases, we can numerically invert $\Gamma_{\k\nu}$.

To further complicate matters, there is also an important interplay between the geometry-statistics duality, and the physical problem at hand which is often overlooked.
If we take the GW problem that we are studying as an example,  we can formally state that the source search and resolution takes place on a n-dimensional
Riemann manifold, with well defined norm, scalar product and metric tensor.  We could initially define all points on this manifold as $\lk\in {\bf R^n}$.  One could further state that 
the physical parameter set is defined by the following subsets : $\{m_1, m_2\}\in[10^2,10^8]\,M_{\odot}$, $z\in[0,20]$, $t_c\in[0,2]$ years, $\{\theta, \iota, \psi\}\in[0, \pi]$ and $\{\phi, \varphi_c\}\in[0,2\pi]$.  To highlight the problem, let us start with the ${\bf R^2}$ sub-space defined by the mass parameters.  As well as the coordinates $(m_1, m_2)$, one could
also construct a sub-manifold by combining any of the two parameters $\{m, \eta, q, M_c, \mu\}$, where $q=m_1/m_2$ and all other quantities have been previously defined.  It is
well known that the coordinates $(m_1, m_2)$ are not a good choice as the parameter space is degenerate, and any map of the form $(m,\eta)\rightarrow(m_1, m_2)$, for 
example, is not a diffeomorphism due to the map not being bijective.  This immediately constrains the choice of ``good" mass parameters.  So, for arguments sake, let's say that
we choose to work with the coordinates $(q, \eta)$, and let us further complicate matters by assuming that we are investigating an almost equal mass binary.  One could 
imagine that our robust detection technique provides solutions of $q=1.1$ and $\eta=0.2494$ for this particular source.  Our subsequent FIM calculation then provides a 
$\pm1\sigma$ error in each of these parameters.  However, for binary systems $q_{min} = 1$ and $\eta_{max}=0.25$, so what exactly does it mean if the FIM predicts error
bound values lower/higher than these values?  In this case, the FIM is blameless.  It assumes a Gaussian distribution of errors and is unaware of any boundaries imposed by
the physical problem at hand.  Geometrically, it assumes that the $(q, \eta)$ sub-manifold is infinite and continuous.  So while there are no geometrical/statistical problems with
this choice of coordinates, astrophysics places a finite boundary on the manifold.  It would therefore seem that for compact binary systems, the best choice of mass parameters
is any combination of $\{m, M_c, \mu\}$, as long as the mapping to $(m_1, m_2)$ space is diffeomorphic.

A final comment involves the structure of the manifold and the physical boundaries on the angular parameters $\{\theta,\phi,\psi,\varphi_c\}$.  We stated earlier that our problem was 
one which could be  formulated in ${\bf R^n}$.  More correctly, due to the sky position coordinates, our manifold actually has a ${\bf S^2\times R^{n-2}}$ structure, i.e. the merging of
a 2-sphere with a $R^{m}$ subspace, where $m=n-2$.  Experience seems to
suggest that the FIM always ``assumes" we are working on an infinite ${\bf R^n}$ manifold with no boundaries.  The fact that the 2-sphere ${\bf S^2}$ is finite may have a knock-on effect on the 
FIM calculation, and may be why the FIM sometimes returns error estimates that are larger than the area of the 2-sphere. Furthermore, if we examine the response function, it should be clear that $h(t)$ is invariant under a translation of the form $(\psi\rightarrow\psi\pm\pi, \varphi_c\rightarrow\pm 2\pi)$, justifying the physical boundaries of $\pi$ and $2\pi$ defined above.  However, at times the FIM can predict $1\sigma$ error estimates in these parameters
that are many multiples of $\pi$ in size.  Common practice would then be to say that the error on these parameters were the size of the physical boundary.  However, it was shown
in Ref~\cite{Cornish2006ms} that doing so causes an underestimation of some of the other parameters.  It was found that, for example, the $(M_c,\varphi_c)$ sub-space is a cylinder embedded 
in the larger parameter space.   On this sub-manifold, the likelihood peak was wrapped multiple times around the cylinder, with a length that was in accordance with the FIM
prediction.

All of this demonstrates that if one is truly intent on using the FIM, one must (a) make sure that the choice of coordinates is optimal, (b) ensure that any mappings between 
the FIM coordinates and true physical parameters are diffeomorphic, (c) make sure that the geometry-statistics duality is fully compatible with the astrophysical implications
, (d) make sure that the condition number is sufficiently small that the results of a numerical inversion are believable and (e) even if everything seems above board, and all
constraints have been respected, be very careful with the interpretation as it may still be incorrect.

%%%%%%%%%%%%%%%%%%%%%%%%%%%%%%%%%%%%%%%%%%%%%%%%%%%%%%%%%%%%%%%%%%%%%%%

\subsection{Markov Chain Monte Carlo}
The goal of any Bayesian analysis is to test a model hypothesis for a particular problem.  For GW astronomy, given a data set $s(t)$, a waveform model based on a set of parameters
$\lk$ and a noise model for $S_n(f)$, one can construct the posterior probability distribution $p(\lk |s)$ via Bayes' theorem
\beq
p(\lk |s) = \frac{\pi(\lk)p(s|\lk)}{p(s)}.
\eeq
The prior probability $\pi(\lk)$ reflects of assumed knowledge of the experiment before we evaluate the data.  $p(s|\lk) = {\mathcal L}(\lk)$ is the likelihood function 
defined by Eq.~\ref{eqn:likelihood}, and $p(s)$ is the marginalized likelihood or ``evidence" given by
\beq
p(s) = \int\, d\lk\, \pi(\lk)p(s|\lk).
\eeq
A common method for carrying out Bayesian analysis is to use a Metropolis-Hastings variant of the MCMC algorithm~\cite{metropolis1953,hastings1970}.  This quite simple algorithm works as follows : starting with a signal $s(t)$ and some initial template $h(t;\lk)$, we choose a starting point $x(\lk)$ randomly within a region of the parameter space bounded
by the prior probabilities $\pi(\lk)$, which we assume to contain the true solution.  We then draw from a proposal distribution and propose a jump to another point in the space $x'(\lnu)$.  To compare the performance of both points, we evaluate the Metropolis-Hastings ratio
\begin{equation}
H = \frac{\pi(x')p(s|x')q(x|x')}{\pi(x)p(s|x)q(x'|x)}.
\end{equation}
Here $q(x|x')$ is a transition kernel that defines the jump proposal distribution, and all other quantities are previously defined.  This jump is then accepted with probability $\alpha = min(1,H)$, otherwise the chain stays at $x(\lk)$.   In order to improve the overall acceptance rate, the most efficient proposal distribution to use for jumps in the parameter space is, in general, a multi-variate Gaussian distribution.  As we assume, in most cases, that the FIM will reliably describe the likelihood surface for the problem at hand, a good starting point is to thus use the FIM.  The multi-variate jumps use a product of normal distributions in each eigendirection of $\Gamma_{\k\n}$.  The standard deviation in each eigendirection is given by $\sigma_{\k} = 1/\sqrt{DE_{\k}}$, where $D$ is the dimensionality of the search space (in this case $D=9$), $E_{\k}$ is the corresponding eigenvalue of $\Gamma_{\k\n}$ and the factor of $1/\sqrt{D}$ ensures an average jump of $\sim 1 \sigma$.  In this case, the jump in each parameter is given by $\delta \lambda^{\k} = {\mathcal N}(0, 1)\sigma_{\k}$. This type of MCMC algorithm is commonly referred to as a Hessian MCMC and is known to have the highest  acceptance rate for this particular family of random walk algorithms.

One problem with the MCMC approach is that, while it provides the full marginalized posterior density function (pdf) for each of the system parameters, it is essentially a black-box algorithm.  The MCMC provides reliable answers provided the chain is run for long enough, and is efficiently sampling the parameter space.   The idea of what
constitutes ``long enough" is both application, and in our situation, source dependent.  Thus MCMC algorithms have to be tailored to the problem in hand.  While there are some defined convergence tests for MCMC algorithms, they are usually tenuous at best.  The best convergence tests are still ``by-eye" tests, such as visually examining the convergence of the means and standard deviations of the chains.  The main issue then becomes ensuring long enough chains to have confidence
in our predictions.  This means that MCMC algorithms take time to map out the pdfs, making a quick estimation out of the question.  

As the convergence of the MCMC can be slow, especially for low mass, high $z$ and/or low SNR systems, we use a mixture model combining a Hessian
and a Differential Evolution (DE) MCMC to estimate our parameters (hereafter DEMC)~\cite{terbraak2006,terbraak2008}.  Differential Evolution is a subset of a genetic algorithm that normally works by 
combining information from
multiple simultaneous chains and using the information to construct jump proposals.  However, the standard implementation of DE can be computationally 
prohibitive, especially for long waveform lengths.  To overcome this problem, we use a  modified version of the DE algorithm.  Once the burn-in part of the chain is
over, we build a trimmed history of the chain, i.e. we keep every 10th chain point.  For each parameter, this then means that we have the full chain array
$x_i$ of instantaneous length $n$,  and a trimmed history array $\bar{x}_i$ with instantaneous length of $\bar{n} = n / 10$.   We then construct our
proposed point in parameter space using
\beq
x_{i+1} = x_i + \gamma \left(\bar{x}_j - \bar{x}_k \right),
\eeq
with $j,k \in U[1,\bar{n}]$ and $i\neq j\neq k$.  The parameter $\gamma$ has an optimal value of $\gamma=2.38/\sqrt{2D}$, where again $D$ is the dimensionality of the 
parameter space.  In this implementation of the algorithm, we use a (2/3, 1/3) split between DE and Hessian MCMC moves.  Furthermore, we know that due to the
shape of the beam pattern functions, the response of an observatory to an impinging GW is naturally bimodal.  To account for this, we set $\gamma=1$ for
10\% of the DE jumps.  This helps moves the chain between modes of the solution and encourages greater exploration.

Finally, whether a move is proposed by the Hessian or DE part of the algorithm, the choice to move or stay requires a calculation of the likelihood.  To accelerate
this calculation and to allow us to investigate the lower mass and/or high redshift systems, we use the composite integral method described in Ref~\cite{Porter:2014sfa} to speed up the 
computational run-time of the algorithm.  To aid practicality for the problem of non-spinning SMBHBs, the parameter set we chose to work with is $\lk = \{\ln(M_{c}), \ln(\mu), \ln(t_{c}), \cos\theta, \phi, \ln(D_{L}), \cos\iota, \varphi_{c}, \psi\}$ where $M_{c}=m\eta^{3/5}$ is the chirp mass, $\mu =m\eta$ is the reduced mass and all other parameters have been previously defined.

%%%%%%%%%%%%%%%%%%%%%%%%%%%%%%%%%%%%%%%%%%%%%%%%%%%%%%%%%%%%%%%%%%%%%%%

\begin{figure}[t!]
 \includegraphics[width=\columnwidth, height=2.5in]{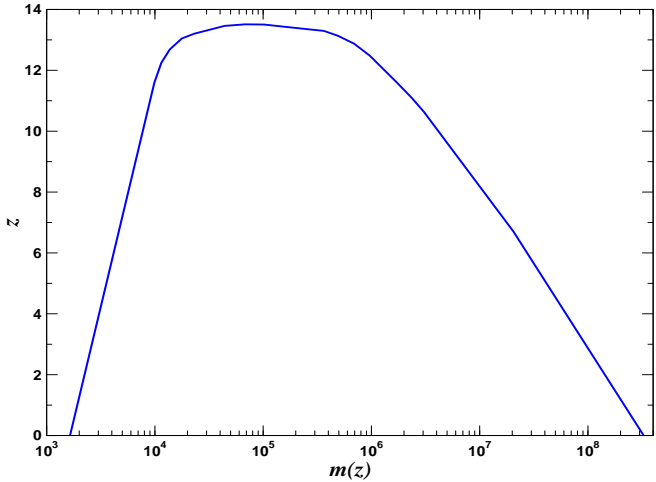}
 \caption{\label{fig:horizon}eLISA redshift horizon distance (excluding merger and ringdown) as a function of redshifted total mass for massive black hole binaries assuming a threshold of $\rho_{th}=10$ on the
 inspiral-only SNR.}
\end{figure}

\section{Astrophysical source selection}\label{sec:astro}
In order to select the specific test-sources for this study, we first generated a population of SMBHBs from a Monte Carlo simulation based on an astrophysical source 
model~\cite{2010ApJ...719..851S}.  To ensure that the comparable mass PN waveform was still valid for the types of sources coming from the population model, we only accepted
sources with mass ratios of $q\in[1,30]$.  Furthermore, to allow us to run long-enough chains in order to have confidence in our Bayesian inference, we restricted all waveforms to an array length of $n\leq 2^{21}$ elements.  This ensured that our Markov chains would run in an acceptable amount of time.  While the mass, mass ratio and redshifts came from the 
distribution, other constraints were imposed by hand : the time to coalescence for each source was randomly chosen between $t_c\in [0.3,1]$ yr and all angular values 
were chosen over their physical ranges.   To choose a detection threshold,  it was recently shown~\cite{Huwyler:2014vva} that a  null test (i.e. the detector 
output consists of instrumental noise only : $s(t) = n(t)$) returns a detection SNR threshold of $\rho_{th} = 10$ for the eLISA configuration, where we define the 
SNR as
\beq
\rho = \frac{\left<s|h\right>}{\sqrt{\left<h|h\right>}}.
\label{eqn:snr}
\eeq
This value 
was again imposed for this study. In Fig~\ref{fig:horizon} we plot the detection horizon $z$ for the sources that returned $\rho \geq \rho_{th}$ as a function of total 
redshifted mass.   We should mention here that the detection horizon is slightly larger to the one presented in Ref~\cite{Huwyler:2014vva} as a more extensive Monte Carlo
was run for this study.
\begin{figure*}
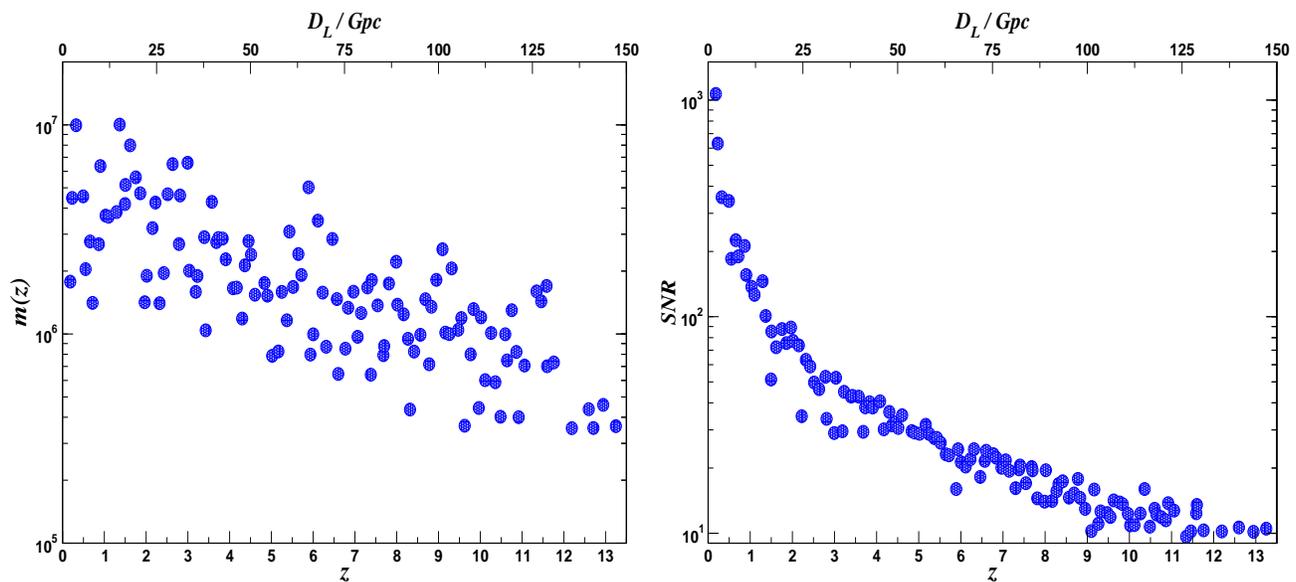

  \vspace{5pt}

  \centerline{\hbox{ \hspace{0.0in} 
    \epsfxsize=2.6in
    \epsfig{file=Mass_vs_Distance.eps, width=3.3in, height=3in}
%   \epsffile{Total_Mass_vs_Distance.eps}
 %   \epsffile{test1month.eps}	
    \hspace{0.05cm}
    \epsfxsize=2.7in
    \epsfig{file=SNR_vs_Distance.eps, width=3.3in, height=3in}
    }
  }

   \caption{ The left hand cell plots the total redshifted mass $m(z)$ of the 120 selected systems as a function of distance. The right hand cell displays  the optimal SNR of each source, also as a function of distance.}
  \label{fig:sources}

\end{figure*}

As the goal of this work is the comparison between different parameter estimation methods, and not the astrophysical significance of source detection and resolution,
we then trimmed the the generated population to the set of study sources as follows : our sources covered a redshift range of  $0.1 \leq z \leq 13.25$.  We then divided this 
range into redshift bins of $\Delta z = 0.1$, and from each bin drew a single source, providing us with 120 sources.  In Fig.~\ref{fig:sources} we plot the total redshifted masses $m(z)$ of the selected
sources as a function of redshift in the left hand panel, and their optimal SNRs as a function of redshift in the right hand panel (i.e. the SNRs obtained by using the
true parameter values).  We can see that from redshifts of $z\geq 6$ onwards, we very quickly begin to approach the detection threshold.  While two of our 
sources fell just below the detection threshold, we decided to study them anyway to investigate if we could detect and resolve the sources.  In these cases, our chains
actually found solutions that differed from the true solution (as would be expected due to the presence of noise), but with SNRs that were superior to the 
detection threshold.  We should point out here that in many of the redshift bins, brighter sources existed.  As a consequence, one should be careful in interpreting 
the results here as an absolute limit of capability for the eLISA detector.  In fact, due to the presence of brighter sources in the redshift bins, for many of the sources presented here,
the results can be interpreted as a ``worst-case" scenario.

Finally, for each source we ran a $10^6$ iteration DEMC, with a ``burn-in" phase of $2\times10^4$ iterations.  During the burn-in phase, we used a combination of 
thermostated and simulated annealing to accelerate the chain mixing and convergence to the global solution~\cite{Cornish2006ms}.  With each chain we checked the instantaneous means 
and standard deviations to ensure they had converged to constant values.  One thing that we observed, was once again how important it is to have long chains to 
properly attain convergence, especially for low mass and/or high redshift sources.  

One further note is that for $z\geq3$, the posterior densities for the vast majority of the sources become at least bi-modal, if not multi-modal.  It has been well documented
that the detector response $h(t)$ is invariant under certain symmetries, either in the polarizations of the waveform, or in the beam pattern functions.  For example, 
a change such as $(\theta,\phi)\rightarrow(\pi-\theta, \phi\pm\pi)$ or $(\theta,\phi, \iota, \psi)\rightarrow(\pi-\theta, \phi\pm\pi, \pi-\iota,\pi-\psi)$ leaves $h(t)$ invariant.
If the signal is strong enough, the Markov chains normally stay on one mode, as it is to unlikely within the run-time of the algorithm to walk across the parameter
space to the other mode(s).  In general, if many chains are run simultaneously, we would expect a 50-50 split between chains ending up on the main mode, and
say, the mode corresponding to the antipodal solution in the sky.  It was clear however, from initial runs, that for systems where the sky error is large, the likelihood 
surface is flat enough that the chain does not have to move very far in parameter space before it can change modes.  In these cases, we ran extended chains and
evaluated results for the dominant mode solution.

We chose to use flat priors in $\{\ln D_L, \ln M_c, \ln\mu, \ln t_c, \cos\iota, \cos\theta, \phi,\psi, \varphi_c\}$, where the boundaries were set at $D_L \in [7.7\times10^{-4},300]$ Gpc,  $M_c \in [435,5.06\times10^7]\,M_{\odot}$, $\mu \in [250, 2.9\times10^7]]\,M_{\odot}$, $t_c\in [0.2, 1.1]$ years,
$\{\cos\iota, \cos\theta, \phi\}\in [-1,1]$ and were left completely open for $\{\psi,\varphi_c\}$.   The prior in luminosity distance is chosen such that we assume that there are no
SMBHBs closer than the M31 galaxy, and no further than a redshift of $z\sim25$.  The priors in $\{M_c,\mu\}$ are chosen so that the minimum total redshifted mass is $m(z)=10^3\,M_{\odot}$, while the upper limit of $m(z)=1.163\times10^8\,M_{\odot}$ assures that the maximum 2PN harmonic last stable orbit frequency corresponds to a 
value of $5\times10^{-5}$ Hz.  These values also correspond to a prior range in reduced mass of $1 \leq q \leq 100$.  The choice of priors that are flat in the logarithm of the dimensionful quantities $\{\ln D_L, \ln M_c, \ln\mu, \ln t_c\}$
says that these parameters are unknown a-priori to within an order of magnitude across a wide prior range. In practice a better approach would be to adopt an astrophysically motivated joint prior
on $\{\ln D_L, \ln M_c, \ln\mu\}$ that takes into account the number of mergers as a function of redshift and the mass evolution of the population. By assigning hyper-parameters to such a prior and
considering the joint posterior distributions for all SMBHBs detected by eLISA, it will be possible to constrain the merger model, in much the same way that the distribution of white dwarf binaries
in our galaxy can be contained by eLISA~\cite{Adams:2012qw}. We leave such a study to future work, but note for now that a uniform prior in $\ln D_L$ corresponds to a $1/D_L$ prior on the
luminosity distance $D_L$, while the FIM study implicitly assumes a uniform prior in $1/D_L$, which corresponds to a $1/D_L^3$ prior on $D_L$. This should be contrasted to a more astrophysically
motivated prior that grows as $D_L^2$ for small $D_L$, before rolling off to some decaying function of $D_L$ for large luminosity distances. Since the results will be dominated by the
likelihood at small $D_L$ and by the prior at large $D_L$, our choice of priors that decay in $D_L$ are not unreasonable.

%%%%%%%%%%%%%%%%%%%%%%%%%%%%%%%%%%%%%%%%%%%%%%%%%%%%%%%%%%%%%%%%%%%%%%%
\begin{figure*}
\vspace{5mm}
\begin{center}
\epsfig{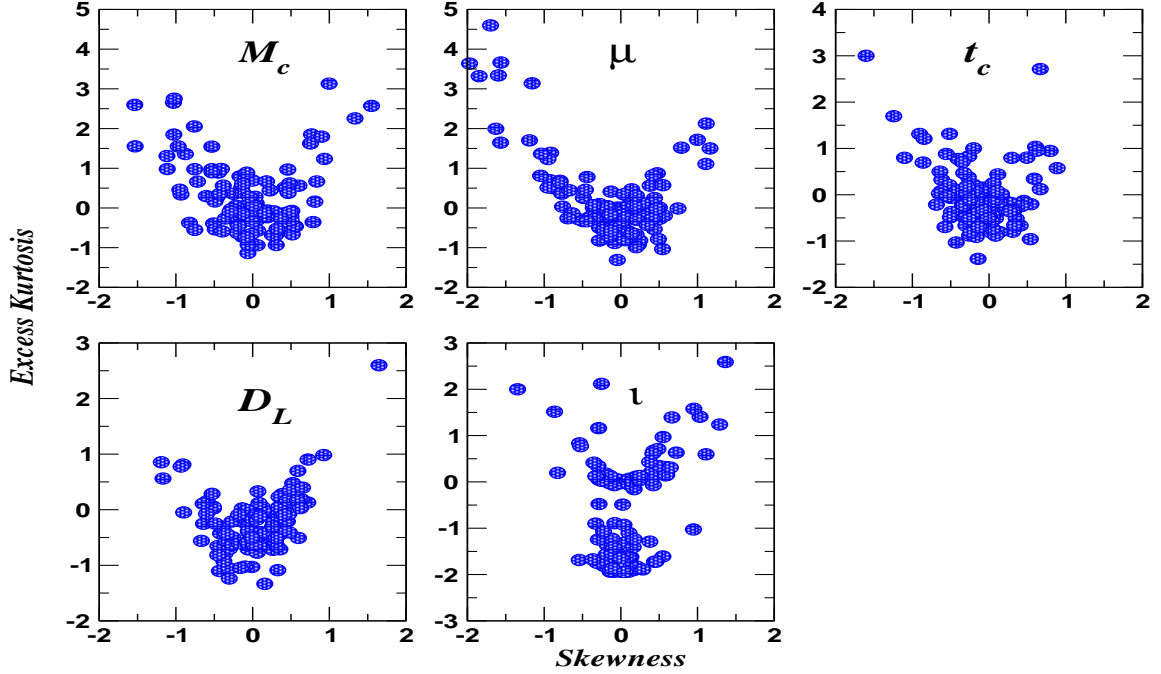}
\end{center}
\caption{A plot of skewness versus excess kurtosis for each of the 120 sources in the study.  The results display that the vast majority of the sources in the study
have non-Gaussian posterior density distributions.}
\label{fig:skewkurt}
\end{figure*}

\section{Results}\label{sec:results}
To compare the FIM and a full Bayesian inference, our plan of action is as follows : the first step is to test the hypothesis of Gaussianity of the parameter posterior 
density distributions, and as a consequence, the validity of using the FIM for parameter estimation.  Once we obtain the levels of Gaussianity, we calculate the
FIM at the median of the distribution, $\tilde{\lambda}^{\kappa}$, for each of the parameters $\{M_c, \mu, t_c, D_L, \iota\}$.  As we are carrying out a Bayesian inference
study, our goal is to estimate the $95\%$ \emph{Bayesian credible intervals} (BCI) for each of the parameters.  For the FIM estimate, we express the $95\%$ quantiles $QI$, as calculated
at $\tilde{\lambda}^{\kappa}$, and assuming a Gaussian distribution in the parameter errors, are simply given by
\beq
QI_{95\%} = \tilde{\lambda}^{\k} \pm 1.96\,\sigma_{\k},
\eeq
where $\sigma_{\k}$ is the standard deviation of the distribution obtained from $\sigma_{\k} = \Gamma_{\k\k}^{-1/2}$.  As shown earlier, Bayes'
theorem can be written in the form
\beq
p\left(\lk | s\right) \propto \pi\left(\lk \right){\mathcal L}\left(\lk \right).
\eeq
Once we have the posterior distribution $p\left(\lk | s\right)$ from the chains, we can now calculate a credible interval ${\mathcal C}$ such that
\beq
\int_{\mathcal C} p\left(\lk | s\right) d\lk = 1-\alpha,
\eeq
where for a $95\%$ BCI, $\alpha = 0.05$.  This allows us to make a degree-of-belief statement that the probability of the true parameter value lying
within the credible interval is $95\%$, i.e.
\beq
{\mathbb P}\left(\lk_{true} \in {\mathcal C} | s \right) = 0.95.
\eeq
For the positional resolution of the source, we can define an error box in the sky according to~\cite{Cutler:1997ta} 
\begin{equation}
 \Delta\Omega = 2\pi \sqrt{\Sigma^{\theta\theta}\Sigma^{\phi\phi}-\left(\Sigma^{\theta\phi}\right)^{2}},
\end{equation}
where 
\begin{eqnarray}
 \Sigma^{\theta\theta} &=& \left<\Delta\cos\theta\Delta\cos\theta\right>,\\
  \Sigma^{\phi\phi} &=& \left<\Delta\phi\Delta\phi\right>,\\
   \Sigma^{\theta\phi} &=& \left<\Delta\cos\theta\Delta\phi\right>,
\end{eqnarray}
and $\Sigma^{\k\n} = \left<\Delta\lambda^{k}\Delta\lambda^{\n}\right>$ are elements of the variance-covariance matrix, found by either inverting the FIM, or are
calculated directly from the chains themselves.

%%%%%%%%%%%%%%%%%%%%%%%%%%%%%%%%%%%%%%%%%%%%%%%%%%%%%%%%%%%%%%%%%%%%%%%

\subsection{Testing the hypothesis of Gaussianity}
The standard assumptions for the validity of using the FIM are : we in the high SNR limit, and we assume that the errors in the estimation of the system parameters are described by a multivariate Gaussian probability distribution 
\beq
p(\Delta \lk) = \sqrt{\frac{|\Gamma|}{2\pi}}\exp\left[ -\frac{1}{2}\Gamma_{\k\nu}\Delta \lk\Delta \lnu\right],
\eeq 
where, once again, we define the FIM
\beq
\Gamma_{\k\nu} = \left<\frac{\partial h}{\partial \lk}\left|\frac{\partial h}{\partial \lnu}\right. \right> = -E\left[ \frac{\partial^2 \ln {\mathcal L}}{\partial \lk \partial \lnu}\right],
\eeq
as the negative expectation value of the Hessian of the log-likelihood, and $|\Gamma| = det(\Gamma_{\k\nu})$.  To test these hypotheses, we calculated the
skewness and kurtosis of the posterior density distributions $p(\lk|s)$.

\begin{figure*}
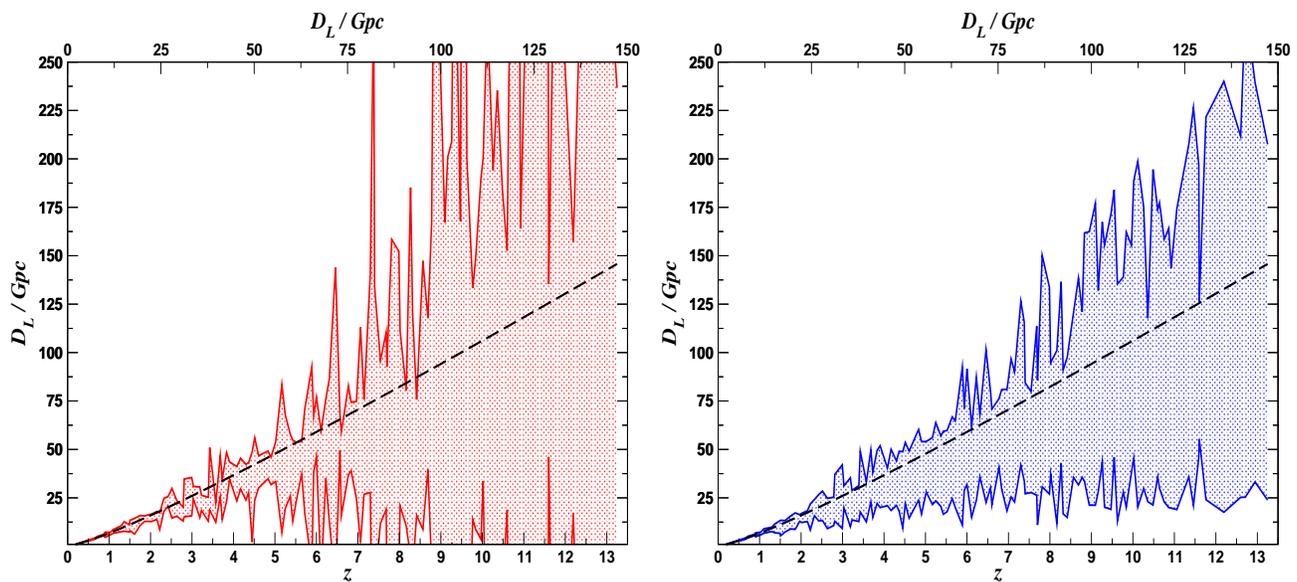

  \vspace{5pt}

  \centerline{\hbox{ \hspace{0.0in} 
    \epsfxsize=2.6in
    \epsfig{file=FIM_Distance.eps, width=3.3in, height=3in}
%   \epsffile{Total_Mass_vs_Distance.eps}
 %   \epsffile{test1month.eps}	
    \hspace{0.05cm}
    \epsfxsize=2.7in
    \epsfig{file=BI_Distance.eps, width=3.3in, height=3in}
    }
  }

   \caption{ In the left hand cell we represent the $95\%$ quantile intervals for the luminosity distance based on a FIM calculation at the chain median.  The
   right hand cell represents the $95\%$ BCI for a Bayesian inference analysis.  Not only do the FIM intervals not contain the true value (represented by the dashed
   line) for some sources, but for many sources at $z\geq 6$ suggest that the source is unresolvable with a $\geq 100\%$ error in the distance estimate.  However, a Bayesian inference shows that the FIM estimation is 
   incorrect and all intervals are not only finite, but importantly, contain the true value.}
  \label{fig:distance}

\end{figure*}

Using the chain points we can now investigate the deviation of each parameter distribution from a normal distribution.  As is standard, we begin with the fact that a random variable $x$ is said to be normal if its density function follows the form
\begin{equation}
f(x) = \frac{1}{\sqrt{2\pi}\sigma}e^{-\frac{1}{2}\left(\frac{x-\mu}{\sigma}\right)^2},
\end{equation}
where $\mu$ and $\sigma$ are the mean and standard deviation respectively.  From this we define the skewness and excess kurtosis via the third and fourth standardised moments
\begin{eqnarray}
\sqrt{\beta_1} &=& \frac{E(x-\mu)^3}{[E(x-\mu)^2]^{3/2}} = \frac{E(x-\mu)^3}{\sigma^3} ,\\ \nonumber \\
\beta_2 &=& \frac{E(x-\mu)^4}{[E(x-\mu)^2]^2} -3 = \frac{E(x-\mu)^3}{\sigma^4}-3,
\end{eqnarray}
where again $E$ denotes the expectation value.  We note here that we are using the excess kurtosis (i.e. with the factor of -3) as for a normal distribution, $\sqrt{\beta_1}= \beta_2\equiv 0$.  For historical reasons, the skewness is defined in this manner even though we can have $\sqrt{\beta_1} < 0$.  For a sample of $n$ data points, i.e. $x_1, .., x_n$, we can write the the sample estimates of $\sqrt{\beta_1}$ and $\beta_2$ in the form 
\begin{eqnarray}
\sqrt{b_1} &=& \frac{m_3}{m_2^{3/2}} \\ \nonumber \\
b_2 &=& \frac{m_4}{m_2^2}-3, 
\end{eqnarray}
where each moment $m_k$ is defined by
\begin{equation}
m_k = \sum_i \left(x_i - \bar{x} \right)^k / n
\end{equation}
and the sample mean $\bar{x}$ is given by 
\beq
\bar{x} = \sum x_i / n.
\eeq
For each of the 120 sources, we analysed the skewness and excess kurtosis of the marginalized posterior densities for the most astrophysically interesting
parameters $\{M_c, \mu, t_c, D_L, \iota \}$.  The results are plotted in Fig.~\ref{fig:skewkurt}.  We remind the reader that if the Gaussian assumption of 
error distribution was true, and hence it was valid to use the FIM, both $\sqrt{\beta_1}$ and $\beta_2$ would be clustered around zero.  It is clear from the 
figure that the vast majority of the marginalized posteriors fail the Gaussianity test for all of the parameters.  This strongly suggests that we should not believe the results of the 
FIM for these sources and should be encouraged to use other methods when carrying out a full parameter estimation for SMBHBs.

\begin{figure*}
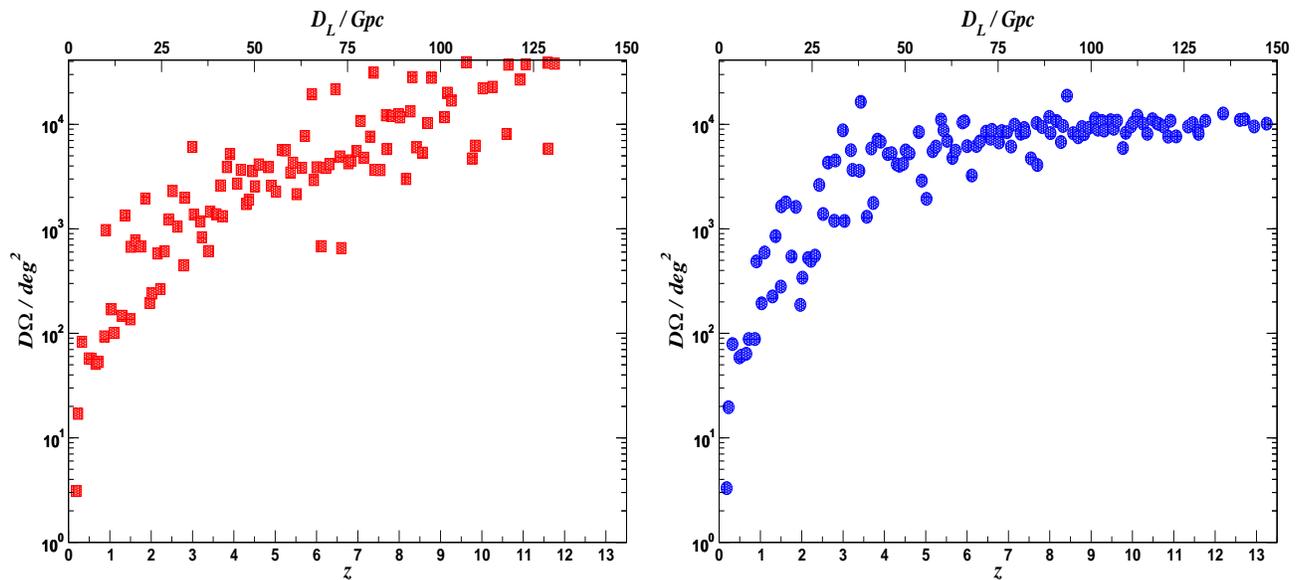

  \vspace{5pt}

  \centerline{\hbox{ \hspace{0.0in} 
    \epsfxsize=2.6in
    \epsfig{file=FisherSkyError.eps, width=3.3in, height=3in}
%   \epsffile{Total_Mass_vs_Distance.eps}
 %   \epsffile{test1month.eps}	
    \hspace{0.05cm}
    \epsfxsize=2.7in
    \epsfig{file=BISkyError.eps, width=3.3in, height=3in}
    }
  }

   \caption{ The left hand cell represents the sky error box in square degrees based on the FIM for the 120 selected systems as a function of distance, while the right hand cell plots the Bayesian inference error box for each source.  The upper limit of each cell corresponds to the total area of the sky.  We can see that for sources at $z\geq 9$, the FIM results suggest that the error in sky location can be equal to or greater than the size of the
   total sky.  However, we can see that from a Bayesian inference study, the results are always finite.}
  \label{fig:sky}

\end{figure*}
%%%%%%%%%%%%%%%%%%%%%%%%%%%%%%%%%%%%%%%%%%%%%%%%%%%%%%%%%%%%%%%%%

\subsection{Luminosity Distance}
As stated in the introduction, the main driving force behind this study are previous works where results based on the FIM suggest that $\Delta D_L/D_L \geq 1$. For 
these systems, the results imply that we would be unable to locate them anywhere in the Universe.  Conceptually, however, this is clearly untrue.  As an example, it 
may be that we detect a source at $z = 10$ with $\rho = 11$.  A FIM analysis returns a value of $\Delta D_L/D_L \geq 1$ for this source, giving a $100\%$ or 
greater error in the distance estimation.  However, if we were able to transport that source from $z=10$ to somewhere in the local group of galaxies, it would have an SNR in the
millions.  The fact that observed SNR is small clearly implies that this source is not that close to us, and in fact, cannot be closer than a certain minimum distance.

In Fig.~\ref{fig:distance} we plot the $95\%$ QI for the luminosity distance based on the FIM estimate in the left hand cell, and the $95\%$ BCI from a Bayesian inference estimate on the right hand
side.  With the FIM results, two things immediately stand out : the first is that for sources at relatively low redshift (i.e. $z\leq 4$), there are some cases where the 
true luminosity distance of the source (represented by the dashed black line) is not within the QI.  The second is that for the majority of sources at $z \geq 6$, the
error in source resolution is $\geq 100\%$.  If we were to stop the analysis here, and accept the results of the FIM, we would conclude that the eLISA observatory,
in its current configuration, is only capable of inferring the distance to sources out to a maximum redshift of $z\sim6$.

On the right hand side of  Fig.~\ref{fig:distance}, the Bayesian inference results yield BCIs with finite limits.  While the BCI naturally grows
in size as we go to higher redshift (due to the SNR approaching the threshold for detection), we can see that for high $z$ sources, we would never claim that they were
closer than $D_L\sim 25$ Gpc (or $z\sim2$).  Furthermore, and contrary to the FIM case, the true result is always contained within the $95\%$ BCI.
We should mention here that there is a bias in the retrieved $D_L$ to lower values than the true values as we go to higher redshift, which we can attribute to the 
results becoming prior dominated at large $D_L$ (low SNR), and our use of prior distributions that scale as inverse powers of $D_L$. This bias would be
removed if we used an astrophysical motivated prior distribution with scale parameters determined from the ensemble of eLISA detections.

One final comment on these results is that, in general, when the BCIs are largest, this corresponds to a source where $m_1\approx m_2$.  As previously stated, in each of the odd
waveform harmonic terms, and in some parts of the higher even terms, we have factors of $\delta m = m_1 - m_2$ appearing.  This means that for the almost equal mass scenario
(which is the case for many of the high redshift sources), a lot of the higher harmonic contributions are suppressed.  This results in a general deterioration in the 
precision of source resolution and an increased error in the parameter estimation.

%%%%%%%%%%%%%%%%%%%%%%%%%%%%%%%%%%%%%%%%%%%%%%%%%%%%%%%%%%%%%%%%%%%

\subsection{Sky Position}
The second major result in this work concerns the error in the location of the source in the sky.  Once again, previous studies based on a FIM analysis have 
presented results where the error box on the sky, $\Delta\Omega$ is equivalent to, or greater than the size of the sky.  While GW detectors/observatories are not
especially good when it comes to sky localisation, one should again expect a finite error box for the sky position.

In Fig.~\ref{fig:sky} we plot the sky error box for the 120 sources using again a FIM analysis (left hand cell) and a Bayesian inference analysis (right hand cell).  In both
cases, the upper limit of the plot corresponds to the full area of the sky.  While not immediately obvious from the plot, the results from both methods rarely coincide.
Furthermore, there is no clear pattern in the results that suggest that one method systematically predicts larger uncertainty regions.  If we focus on the FIM error box, we 
can see that at $z\geq9$ we start to observe sources where the error box contains the entire sky, and in fact after $z\geq 11$ nearly all sources in the sample 
have FIM error boxes that are greater than the totality of the sky.  On the other hand, the Bayesian results are always finite.  While the errors grow quite quickly
as a function of redshift, the maximum error box seems to converge to 1/4 of the total sky for distant sources.  

We should once again remind the reader that these results reflect the nature of the sources in our sample and should not be
taken as an absolute reflection of the capabilities of the eLISA detector.  While we do not expect the high redshift behaviour to change, nature may be kind enough
to give us sources at lower redshift that have much smaller error boxes, and allow a coincident analysis with EM telescopes.

%%%%%%%%%%%%%%%%%%%%%%%%%%%%%%%%%%%%%%%%%%%%%%%%%%%%%%%%%%%%%%%%%%%%%

\subsection{System Parameters}
\begin{figure*}
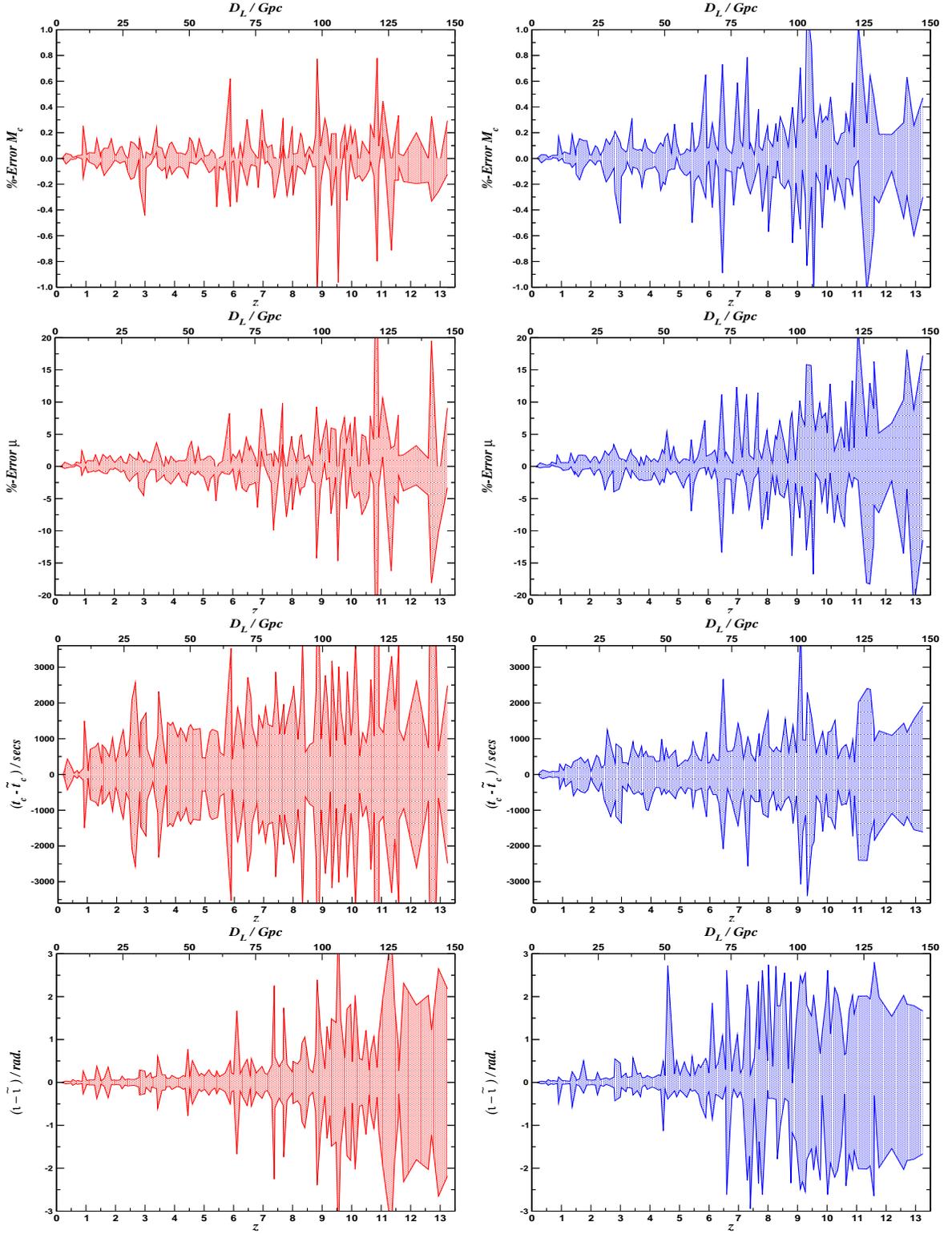

\begin{center}
  \vspace{5pt}

  \centerline{\hbox{ \hspace{0.0in} 
    \epsfxsize=2.6in
    \epsfig{file=McFCI.eps, width=3.in, height=2.in}
    \hspace{0.05cm}
    \epsfxsize=2.7in
    \epsfig{file=McBCI.eps, width=3.in, height=2.in}
    }
  }
  \centerline{\hbox{ \hspace{0.0in} 
    \epsfxsize=2.6in
    \epsfig{file=muFCI.eps, width=3.in, height=2.in}
    \hspace{0.05cm}
    \epsfxsize=2.7in
    \epsfig{file=muBCI.eps, width=3.in, height=2.in}
    }
  }
  \centerline{\hbox{ \hspace{0.0in} 
    \epsfxsize=2.6in
    \epsfig{file=tcFCI.eps, width=3.in, height=2.in}
    \hspace{0.05cm}
    \epsfxsize=2.7in
    \epsfig{file=tcBCI.eps, width=3.in, height=2.in}
    }
  }
  \centerline{\hbox{ \hspace{0.0in} 
    \epsfxsize=2.6in
    \epsfig{file=incFCI.eps, width=3.in, height=2.in}
    \hspace{0.05cm}
    \epsfxsize=2.7in
    \epsfig{file=incBCI.eps, width=3.in, height=2.in}
    }
  }

   \caption{ $95\%$ quantile (left hand side) and credible (right hand side) intervals for $M_c$ (row 1), $\mu$ (row 2), $t_c$ (row 3) and $\iota$ (row 4).  The intervals are expressed
   as a percentage error for the mass parameters, while the boundaries of row 3 correspond
   to an error of $\pm1$ hour in the time of coalescence.  As we know that the FIM estimates of errors in $D_L$ and $\Delta\Omega$ can be greater than $100\%$ at redshifts of
   $z\geq6$, we should not believe any of the results in the left hand column beyond this redshift as they are clearly incorrect also.}
  \label{fig:parameters}
\end{center}
\end{figure*}

In Fig.~\ref{fig:parameters} we plot the $95\%$ QIs and BCIs as a percentage error for the two mass parameters $M_c$ (row 1) and $\mu$ (row 2), and the intervals centered around the median subtracted true values (i.e. $\lambda_{true}^{\k} - \tilde{\lambda}^{\kappa}$) for $t_c$ (row 3) and $\iota$ (row 4), for each of the 120 sources.  In all cases, the column on the left hand side represents the FIM estimate, 
while the Bayesian inference estimate is represented by the column on the right.  To ease interpretation, the y-axis for the time of coalescence is given in seconds (with the upper and lower boundaries of the plot corresponding to $\pm1$ hour), and
the y-axis for inclination is given in radians.  We remind the reader that the Bayesian inference results are the ``gold standard" results, and should be interpreted as
being more exact.  Finally, we point out here that the median value in all cases lies within the BCIs.

If we first focus on the results for the two mass parameters : in both cases, the estimations are quite remarkable.  For the FIM analysis, the percentage error in the chirp mass is
always less than $\pm1\%$, and less than $\pm10\%$ for the reduced mass in the vast majority of cases.  It is only as we approach redshifts of $z\sim11$ that we see the maximum
error in the reduced mass of $\pm22\%$.  The Bayesian inference results paint a similar picture with the vast majority of cases returning a BCI of less than $\pm1\%$ for the 
chirp mass, and $\pm10\%$ for the reduced mass.  Once again, it is only at high redshifts that we attain maximum errors of $\pm1.2\%$ in $M_c$ and $\pm20\%$ for $\mu$.  
While there does not seem to be much difference in both methods, we should point out that in the FIM analysis, most of the the sources at redshifts of $z\geq6$ have a 
$100\%$ error in $D_L$ and/or $\Delta\Omega$.  In fact, by referring to these parameters, we should not trust any of the FIM results at redshifts of $z\gtrsim6$.   We should once again remind 
the reader, in order to demonstrate just how good these results are,  that the best observed candidate systems for SMBHBs are at low redshift ($z\leq 0.4$), yet still 
have order of magnitude errors in the estimation of the mass parameters.

In row 3, we compare the estimates for the precision in the prediction of the time of coalescence.  A remarkable result of this study is that the maximum Bayesian credible
interval is $\pm 1$ hour at $z\sim 9$.  For the vast majority of sources, the predictive power of when these systems will coalesce is on the order of minutes.  Finally,
in row 4 we present the BCIs for the inclination of the source.  As inclination is important in the estimation of the individual masses for an electromagnetically 
observed binary, we can see that it should be possible to make reasonable estimations of the inclination of the source out to redshifts of $z\sim5$, with a maximum
error in the prediction of the inclination of around 60 degrees.

%%%%%%%%%%%%%%%%%%%%%%%%%%%%%%%%%%%%%%%%%%%%%%%%%%%%%%%%%%%%%%%%%%%%%%%

\section{Conclusion}
We have compared the error predictions from both a Fisher information matrix and a Bayesian inference approach for 120 SMBHB sources at redshifts of 
between $0.1\leq z\leq 13.2$.  As has been observed in previous parameter estimation studies, the FIM once again predicts that for certain sources, the error in the
estimation of luminosity distance and the position of the source in the sky is $100\%$.   Taking the FIM results on face value, one would conclude that the eLISA 
detector in its current configuration is only capable of source resolution out to a redshift of $z\sim 6$.  However, a full Bayesian inference analysis shows this
conclusion to be false.  The error predictions for the luminosity distance are not only always finite, but for each individual source set a minimum distance
below which we would never mistake the source as being.  It is well known that GW observatories have poor angular resolution, and while the error box on the sky grows
as a function of redshift, this error box is always a finite quantity.  While not the main goal of this work, we
also demonstrate that out to redshifts of $13.2$, we should be able to predict the chirp mass with a maximum error of $\lesssim 1\%$, the reduced mass with a maximum
error of $\lesssim20\%$, the time to coalescence with a maximum error of 2 hours, and to a redshift of $z\sim5$, the inclination of the source with a maximum error of 
$\sim60$ degrees.

Our study also demonstrated that the posterior distributions for the parameters describing SMBHBs are highly non-Gaussian.  This should make one
highly suspicious of any FIM result as one of the principal condition hypotheses for usage of the FIM is clearly violated. The FIM will continue to be a useful
tool when one is searching for a quick ``order of magnitude" estimate of parameter errors.  However, we hope that this study has clearly shown that for 
full parameter estimation studies, and especially for mission configuration science impact studies, other more robust methods are needed.

\bibliography{bibliography}

\end{document}